\documentclass[apj]{emulateapj}

\shorttitle{Kinematics of Ionized Gas around TW Hya}
\shortauthors{Goto et al.} 

\begin{document}

\title{Kinematics of Ionized Gas at 0.01~AU of TW~Hya\altaffilmark{1}}

\author{M. Goto\altaffilmark{2},
A. Carmona\altaffilmark{3},
H. Linz\altaffilmark{2},
B. Stecklum\altaffilmark{4},
Th. Henning\altaffilmark{2},
G. Meeus\altaffilmark{5,6},
T. Usuda\altaffilmark{7}}

\altaffiltext{1}{Based on data collected by SINFONI observations
                 [79.C-0559(B)] at the VLT on Cerro Paranal (Chile),
                 which is operated by the European Southern
                 Observatory (ESO).}

\altaffiltext{2}{Max-Planck-Institut f\"ur Astronomie, 
                 K\"onigstuhl 17, D-69117 Heidelberg, Germany}

\email{mgoto@mpia.de}

\altaffiltext{3}{ISDC, Ch. d'Ecogia 16, CH-1290 Versoix, Switzerland,
  and  Observatoire de Gen\`eve, University of Geneva, Ch. des
  Maillettes 51, 1290 Versoix, Switzerland}

\altaffiltext{4}{Th\"uringer Landessternwarte Tautenburg, Sternwarte 5,
                 D-07778 Tautenburg, Germany}

\altaffiltext{5}{Astrophysikalisches Institut Potsdam, An der
  Sternwarte 16, 14482 Potsdam, Germany}  

\altaffiltext{6}{Universidad Aut\'onoma de Madrid, Departamento
  de F\'isica Te\'orica, Cantoblanco, 28049 Madrid, Spain}

\altaffiltext{7}{Subaru Telescope, 650 North A`ohoku Place,
                 Hilo, HI 96720, USA}

\begin{abstract}

  We report two-dimensional spectroastrometry of Br$\gamma$
  emission of TW~Hya to study the kinematics of the ionized gas
  in the star--disk interface region. The spectroastrometry with
  the integral field spectrograph SINFONI at the Very Large
  Telescope is sensitive to the positional offset of the line
  emission down to the physical scale of the stellar diameter
  ($\sim$0.01~AU). The centroid of Br$\gamma$ emission is
  displaced to the north with respect to the central star at the
  blue side of the emission line, and to the south at the red
  side. The major axis of the centroid motion is P.A.$=
  -$20\degr, which is nearly equal to the major axis of the
  protoplanetary disk projected on the sky, previously reported
  by CO sub millimeter spectroscopy (P.A.$= -$27\degr)
The line-of-sight motion of the Br$\gamma$
  emission, in which the northern side of the disk is
  approaching toward us, is also consistent with the direction
  of the disk rotation known from the CO observation.  The
  agreement implies that the kinematics of Br$\gamma$ emission
  is accounted for by the ionized gas in the inner edge of the
  disk. A simple modeling of the astrometry, however, indicates
  that the accretion inflow similarly well reproduces the
  centroid displacements of Br$\gamma$, but only if the position
  angles of the centroid motion and the projected disk ellipse
  is a chance coincidence. No clear evidence of disk wind is
  found.
\end{abstract}

\keywords{circumstellar matter --- planetary systems:
  protoplanetary disks --- stars: activity --- stars: formation
  --- stars: individual (TW~Hya) --- stars: pre-main sequence
  --- techniques: high angular resolution}


\section{Introduction}

It is being commonly accepted that a protoplanetary disk
dissipates inside out. Although the direct trigger is still not
clear, a large number of spectral energy distributions collected
by the {\it Spitzer Space Telescope} \citep[e.g.,
][]{cur08,sic08}, as well as the direct imaging of the central
cavity by the submillimeter/millimeter interferometric arrays
(submillimeter array (SMA): Brown et al. 2008; Plateau de Bure
Interferometer: Pi\'etu et al. 2006), give clear observational
supports that many disks dissipate at the central part
first. The innermost region of a disk ($<$ 1~AU) is, therefore,
the critical site to study how the disk dissipation starts. The
structure of a disk at that radius is, however, expected to be
complicated \citep[see ][ for a recent review]{dul10}. Dust
grains evaporate in close proximity to the star. Hot gas may
remain inside the inner truncation of the dust disk, but in
ongoing dissipation either by accretion onto the star, disk
wind/jet, or by a forming planet that sweeps out the disk.

TW~Hya belongs to a small group of stars that the disk
dissipation is currently in progress \citep[][ for a review of
transition objects]{ale07}. Close vicinity of TW~Hya to the
solar system \citep[50--60~pc; ][]{kas97,hof98,mam05} makes it
an exceptional laboratory to look into the central few AU of a
disk (we will use $d=$51~pc from \citet{mam05} throughout this
paper). Missing inner disk of TW~Hya was first recognized by
\citet{cal02} through the lack of near-infrared excess, which
attests the absence of hot dust close to the central star. A
comprehensive multiwavelength study of the mass distribution of
the disk by \citet{gor11} confirmed that the inner disk
($<3.5$~AU) is indeed depleted by one to two orders of magnitude
than that is extrapolated from the surface density of the disk
immediately outside.  Interferometric observations also lend
support to the clearing of the inner disk, although the exact
location of the inner truncation is under debate
\citep[0.4--4~AU; ][]{eis06,rat07,hug07}.

\begin{figure*}
\begin{center}
\includegraphics[height=0.6\textheight,angle=-90]{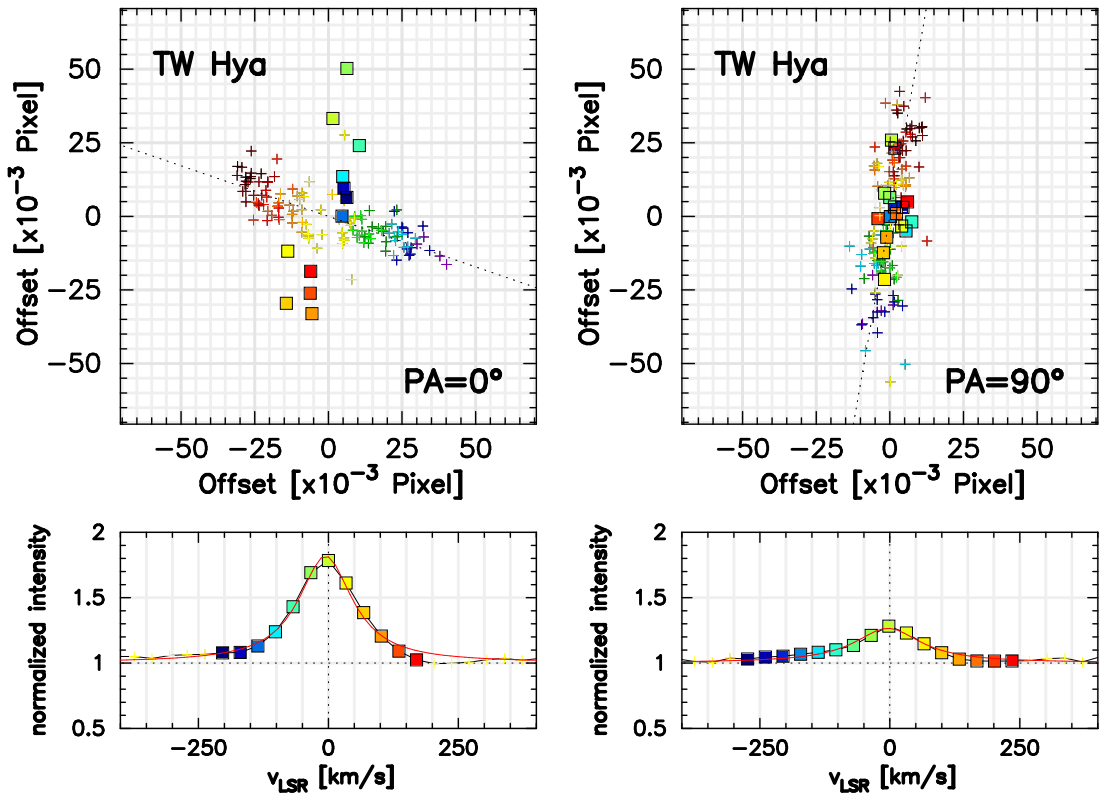}
\end{center}
\caption{Top: the centroids of the PSF at the wavelengths near
  Br$\gamma$ emission. North is up, and east is to the left.
  The centroids recorded with the detector column aligned to
  P.A.=0\degr~are shown in the left panel, and to
  P.A.=90\degr~in the right panel. The PSF centroids in the
  Br$\gamma$ emission are shown with squares, and those of the
  continuum wavelengths are shown with crosses. Both symbols are
  color-coded from blue to red with the wavelength. The drifts of
  the PSF at the continuum wavelengths are fitted by linear
  functions of the wavelength, and are shown in dotted
  lines. Bottom: line profiles of the Br$\gamma$ emission. The
  crosses and the squares are from the observations, color-coded
  in the same way with the panels above. The red lines are
  Lorentzian functions that best fit the observed data points.
  \label{f1}}
\end{figure*}

Hydrogen recombination lines are widely used to study the hot
plasma near the central star, in particular in the accretion
columns channeled by the magnetic fields
\citep{uch84,cam90,koe91}. However, the quantitative modeling of
the line emission is still a challenge
\citep{rom04,gre06,kur06,kur08}, and the origin of the emission
line is not completely understood, either \citep[e.g.,
][]{kra08}. The goal of this paper is to study the dynamics of
the ionized gas inside the inner hole of a transition disk with
a high angular precision afforded by two-dimensional (2D)
spectroastrometry \citep[e.g., ][]{gar99,dav10}.
Spectroastrometry with long slit spectrograph has been used for
some time to study the kinematics of the emission lines of
pre-main-sequence stars on the order of 1~AU
\citep[e.g., ][]{sol93,bai98}. We will employ an integral field
spectrograph assisted by an adaptive optics system to achieve
the astrometry of the physical scale of a stellar diameter in
order to deal with the ionized gas in the star--disk interface
region.

\section{Observation}

The observations were performed by the integral field
spectrograph SINFONI \citep{eis03} at the Very Large Telescope
(VLT) with the adaptive optics system MACAO \citep{bon04}. An
image on the focal plane (field of view 0\farcs8 $\times$
0\farcs8) was sliced into 32 reflective slitlets, each providing
a medium resolution spectrum in the $K$ band ($R=$4000). The
pixel scale is 12.5~mas on the detector, although the spatial
sampling perpendicular to the length of the slitlets is limited
by the resolution of the image slicer (25~mas). The observations
were carried out with two instrumental angles, one with the
detector column pointing to the north (P.A.$=$0\degr; performed
on UT 2006 January 19), and the other to the east (P.A.$=$90\degr;
on UT 2006 January 28).

\begin{figure*}
\begin{center}
\includegraphics[height=0.6\textheight,angle=-90]{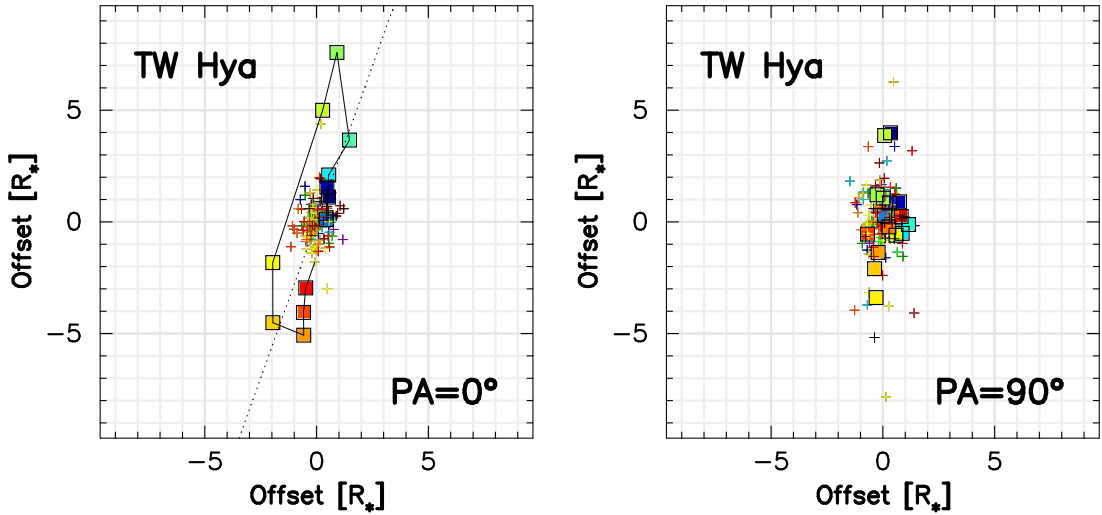}
\end{center}
\caption{Same as in Figure~\ref{f1}, but after the correction for
  the PSF drift. The offsets of the PSF centroids are converted
  to the physical scale, assuming the distance to TW~Hya is
  51~pc \citep{mam05}. The major axis of the locus is oriented
  to P.A.=$-$20\degr~(left). \label{f2}}
\end{figure*}

\section{Data Reduction}

\subsection{Spectroscopy}

The SINFONI data-reduction pipeline provides data cubes
  that the monochromatic images are stacked along the
  wavelength. One-dimensional (1D) spectra were extracted from the
data cubes by integrating the flux inside the aperture of
0\farcs2. The spectroscopic standard star HD~37702 (B8~V) was
used to remove the telluric lines. The photospheric absorption
line of HD~37702 at Br$\gamma$ was removed before dividing the
object spectra by subtracting a Lorentzian function fitting the
absorption line profile. The wavelength calibration was
performed by cross-correlating the object spectra with the
atmospheric transmission curve calculated by ATRAN
\citep{lor92}.  The error of the wavelength calibration was
  evaluated against the photospheric absorption lines at
  2.10--2.21~$\mu$m (\ion{Al}{1}, \ion{Mg}{1}, \ion{Na}{1}), and
  found to be accurate to 20~km~s$^{-1}$. The observed line
velocity was converted to the velocity with respect to the local
standard of rest, using {\it rv} package on IRAF.\footnote{IRAF
  is distributed by the National Optical Astronomy
  Observatory, which is operated by the Association of
  Universities for Research in Astronomy, Inc., under
  cooperative agreement with the National Science Foundation.}
The equivalent width of the line emission was reduced to
$\sim30$\% of that on 2006 January 19 (P.A.=0\degr), when we
observed the object again on 2006 January 28 with the instrumental
position angle 90\degr.  The variability of Br$\gamma$ emission
of a similar amplitude has been reported by \citet{eis11} for
TW~Hya.

\subsection{Drift Correction}

The point-spread function (PSF) of the star near Br$\gamma$
emission was fit with a 2D Gaussian function at each wavelength
to measure the PSF centroid. The positions of the PSF centroid
have shown an overall drift along P.A.$=$71\degr~with the
wavelength in the case of the instrumental angle aligned to
0\degr (crosses in Figure~\ref{f1}, left).  This is apparently
an instrumental effect, left uncorrected by the image alignment
process in the pipeline, as the drift orientation changes by
about the same amount when the instrument is rotated by
90\degr~(the drift P.A.$= -$10\degr, Figure~\ref{f1},
right). The overall motion of the PSF drift was fit by a linear
function of the wavelength and subtracted from the
measurements. The centroid position at Br$\gamma$ emission after
the correction is shown in Figure~\ref{f2}.  The motion of the
emission centroid of Br$\gamma$ was unfortunately aligned to the
orientation of the PSF drift in the data with the instrumental
angle set to P.A.$=$90\degr. Although the results of the two
observations performed with different rotation angles are
qualitatively consistent, we will use the data from
P.A.$=$0\degr~ mostly in the following discussion, as it is less
affected by the correction of the PSF drift. The locus of the
centroid motion is an elongated ellipse with the major axis
aligned to P.A.$=-$20\degr~in the data obtained with
P.A.$=$0\degr.  The total extent of the loop is about
0.05~pixel, or 0.04~AU (8$R_\ast$) from the central star at the
distance of TW~Hya \citep[$R_\ast$=0.96~$R_\odot$; ][]{yan05}.

\begin{figure*}
\begin{center}
\includegraphics[height=0.75\textwidth,angle=-90]{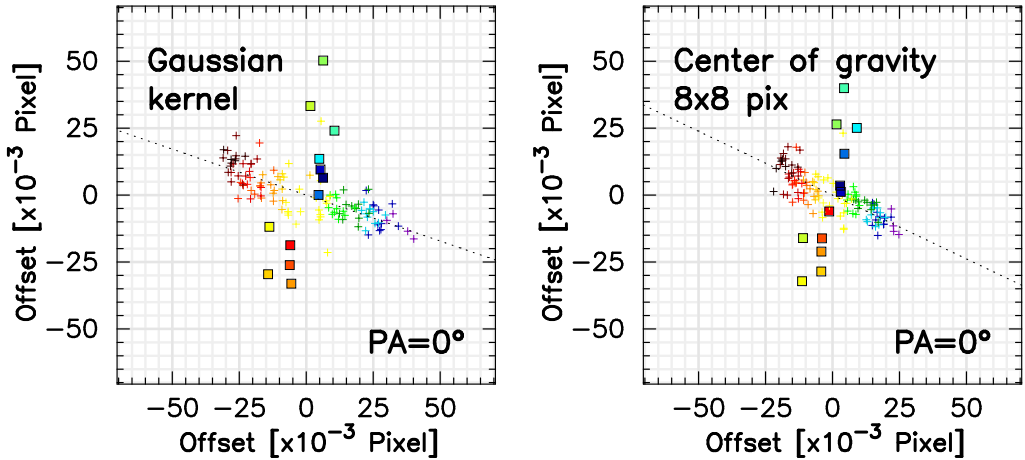}
\end{center}
\caption{Comparison of the astrometry measured by Gaussian
  fitting (left) to that by the center of gravity (right).  The
  center of gravity (flux-weighted center of the emission) is
  calculated within 8$\times$8 pixel windows centered at the
  peak of the PSF.  Two measurements are consistent, attesting
  the positive astrometric signal at Br$\gamma$ emission is not
  an artifact of the measuring scheme. \label{f3}}
\end{figure*}

\subsection{Astrometric Accuracy}

The smallest centroid offset that we can detect in the
observation was estimated from the dispersion of the measured
centroid positions at the continuum wavelengths. The standard
deviation of the amplitudes of the centroid positions at the
continuum (Figure~\ref{f2}) is $5\times10^{-3}$ pixel, which is
0.003~AU at the distance of TW~Hya, or less than one stellar
radius. We will check below if the high astrometric accuracy
achieved is consistent with the simple photon statistics. In the
case that the error distribution is canonical, the accuracy of
the astrometry improves from the size of PSF proportionally to
the squared root of the total photons detected \citep[e.g.,
][]{bai98}. The size of the PSF during the observation of TW~Hya
measured in the median frame of the image stack was 65~mas in
the full width at half-maximum (FWHM). The integrated flux of
the star in a raw frame was $\sim1\times10^5$~ADU for 30\,s of
exposure time at the continuum wavelengths near Br$\gamma$
emission. The total exposure time was 360\,s for the object for
one instrumental rotation. The photon count of the fully
processed image we used for the astrometry is therefore
$\sim2.9\times10^6$ per spectral unit, after converting ADU to
photons by the detector gain 2.4~ADU/electron.  The nominal
statistical error of the astrometry is therefore 0.038~mas or
3.1$\times 10^{-3}$~pixel. The astrometric accuracy achieved is
about $1\times10^{-2}$ pixel in terms of the FWHM, therefore
consistent with the photon statistics, though it is a few times
worse than the ideal case.

We can calculate a formal error of the center of gravity for
more general cases, using the realistic instrument parameters of
SINFONI.  The flux-weighted centroid of an emission source is
given by
\[
z(\lambda) = \frac{\sum\limits_i{x_i  y_i}}{\sum\limits_i{y_i}},
\] 
where $x_i$ is the 1D coordinate of the detector either row or
column, and $y_i$ is the photon count of the pixel $x_i$. The
formal error of the centroid position $\sigma_z$ is given by the
pixel counts and their errors $\sigma_{y_i}$,
\[
\sigma_{z}^2 = \sum\limits_i  \left[\frac{ x_i \sum\limits_i y_i- \sum\limits_i x_i y_i }{(\sum\limits_i y_i)^2} \cdot \sigma_{y_i} \right]^2.
\]
We calculated $\sigma_{y_i}$ as a squared sum of the readout noise
and the shot noise of the pixel, 
\[
\sigma_{y_i}^2  = (\sigma_{\rm r} ^2 + y_i ) \cdot n_{\rm fr},
\] 
where $n_{\rm fr}$ is the number of frames averaged. The readout
noise $\sigma_{\rm r}$ is about $10$~electron for a single
readout for SINFONI. The pixel count $y_i$ is evaluated with
actual observed data in the unit of electrons, and before the
subtraction of the sky emission; although the sky background is
negligible at the wavelength concerned (0--10~ADU). The
formal error $\sigma_z$ is a function of the size of the area in
which the summation is calculated and is smallest when the
minimum number of the pixels is involved. $\sigma_z$ is on the
order of 0.01~pixel from 4$\times$4 to 16$\times$16 pixel
windows, which is in reasonable agreement with 0.005~pixel that
is actually achieved. The center of gravity measured in
8$\times$8 pixel window is compared with the measurements of the
centroid by 2D Gaussian fitting and found to be consistent
(Figure~\ref{f3}).

\subsection{Test Against Artifact}

\begin{figure*}
\begin{center}
\includegraphics[height=1.0\textwidth,angle=-90]{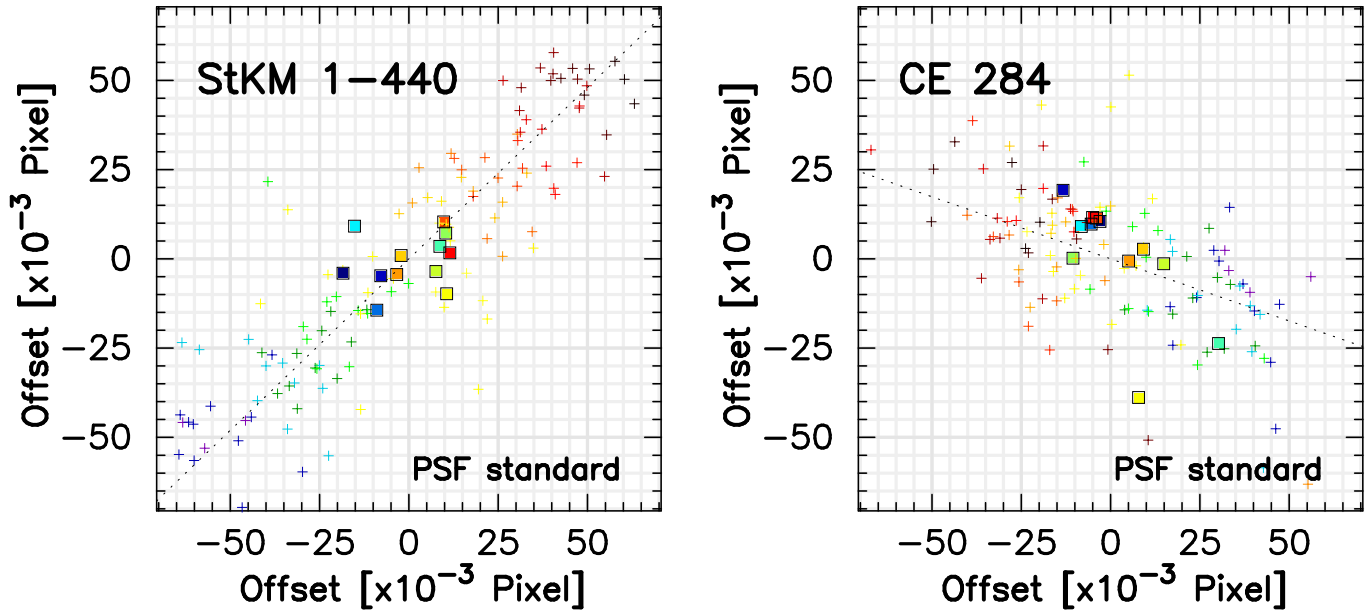}
\end{center}
\caption{Comparison of the astrometry of TW~Hya with
  P.A.=0\degr~ to that of two PSF standard stars StKM~1-440 and
  CE~284 at the same wavelength with Br$\gamma$ emission.  Note
  that the astrometric accuracy of the PSF stars is not as good
  as TW~Hya, because the photon statistic is worse.
  Nevertheless, it is reasonably clear that no positive astrometric signal
  is detected in either of the PSF standard stars, if the
  amplitude of the possible false signal is as large as
  TW~Hya.  \label{f4}}
\end{figure*}

A detection of a spectroastrometric signal is often tested with
photospheric emission lines, in which little or no offset of the
emission centroid is expected. Unfortunately, we cannot
  afford such a test, because Br$\gamma$ emission is the only
  significant emission line in our spectral coverage.  Instead,
we measured the astrometry of PSF standard stars at
  the wavelength of Br$\gamma$ line with the same method
that we used for TW~Hya in order to show that the offset
of the centroid detected is not an artifact of the measuring
scheme. The PSF standard stars StKM~1-440 and CE~284 were
observed on UT 2006 January 24 and 28, respectively. They are
  fainter than TW~Hya ($K$=7.297~mag; StKM~1-440 $K$=8.98~mag;
  CE~284 $K$=9.39~mag) and were observed with less telescope
time. The astrometric accuracy achieved was therefore a
few times worse, as the photon statistics are not as good as
TW~Hya.  Nevertheless, the astrometric accuracy is good enough
to negate the possible offset of the centroid at the wavelength
of Br$\gamma$, in the sense that it should have been detected,
if the astrometric signal in TW~Hya were an artifact and were
present in the data of StKM~1-440 and CE~284 with the same
amplitudes (Figure~\ref{f4}).

The advantage of 2D spectroastrometry is not only a better
spatial coverage and better photon statistics, but also the
robustness against the artifacts. The most of the artifacts that
spectroastrometry suffers from, come from the use of a slit and
vignetting of the source, combined with uneven illumination of a
slit and a distorted PSF \citep{bra06}.  In 2D
spectroastrometry, we do not use a slit but slitlets, where the
problem of vignetting is expected to be much smaller.

\subsection{Convolution with Continuum Emission}

Note that the displacement of the centroid of Br$\gamma$
  emission we measured above is a lower limit of the real offset
  of the emission line, because the PSF of the line emission is
convolved with that of the continuum emission.  The
  convolution shifts the line emission centroid toward the
  continuum emission source, if it is measured by fitting a
  Gaussian function.  If we assume that the amount of the
shift is proportional to the flux ratio, the convolution
  moves the observed centroid to
  \[r_{\rm obs} = \frac{r_{\rm cont} F_{\rm cont} + r_{\rm line}
    F_{\rm line}}{F_{\rm cont} + F_{\rm line}},\] where $r_{\rm
    line}$ is the intrinsic location of the centroid of the
  Br$\gamma$ emission before the convolution, and $F_{\rm cont}$
  and $F_{\rm line}$ are the flux of the continuum and the
  continuum-subtracted line emission, respectively. Effect of
  such convolution can be seen in Figure~\ref{f1}, where the
  amplitude of the centroid motion is smaller in the data with
  the instrumental angle set to P.A.$=$90\degr~than that to
  P.A.$=$0\degr. This is reasonable, since the line intensity
  with respect to the continuum emission is three times smaller
  in P.A.$=$90\degr.  As we are only interested in the location
  of the line centroid relative to that of the continuum
  emission which is supposed to be the position of the star, we
  can set $r_{\rm cont} = 0$ and $r_{\rm line}$ is given by
\[r_{\rm line} = r_{\rm obs} (1 + F_{\rm cont} / F_{\rm
  line}).\] Such a simple correction, though frequently
  used, is problematic, because the correction factor $1 +
F_{\rm cont} / F_{\rm line}$ is large at the line wings where
the line emission is weak compared to the continuum
emission ($F_{\rm line} / F_{\rm cont} \approx 0$); and the
  astrometric signal, as well as its error, diverge into
  infinity at the continuum wavelengths.\footnote{In our
  case, the line-to-continuum ratio at the line center is 0.75
  (P.A.=0\degr) and 0.25 (P.A.=90\degr), therefore the
  correction factor is at minimum $1 + F_{\rm cont} / F_{\rm
    line} >2.3$ and $>5$. The correction factor at the line
  wings is much larger than the line center.} To avoid that,
one could apply the correction only at the wavelengths where the
line emission is detected above a certain significance
($F_{\rm line} / F_{\rm cont} \gg 0$). However, such a
  selective correction has a pitfall to distort the geometry of
  astrometry by exaggerating astrometric signals at the border
  of the wavelength interval that the corrections were applied.

  For these reasons, we choose not to apply the correction
    for the continuum convolution with the understanding that we
    will refrain from the quantitative discussion concerning the
    physical scale and the complex aspects of the geometry of
    the ionized gas.  We instead focus on the position angle of
    the displacement, which is more critical for the
    interpretation of the astrometric signal and is less
    affected by the convolution with the continuum emission.


\begin{table}
\begin{center}
  \caption{Geometrical Parameters of TW~Hya Stellar--Disk System
   Assumed Throughout the paper.}\label{tb1} 
  \begin{tabular}{lll}
\hline \hline
Parameter                         &   Values              &  Reference \\

\hline

Stellar mass              & 0.8~$M_\odot$      &  \citet{cal02}  \\
Stellar radius            & 0.96~$R_\odot$     &  \citet{yan05}  \\
Co-rotational radius      & 6.3~$R_\ast$       &  \citet{joh01}  \\
Disk P.A.\tablenotemark{a}& $-$20\degr         & Present observation \\ 
Disk inclination angle    & 7\degr             &  \citet{qi04,qi06}  \\
Distance                  & 51~pc              &  \citet{mam05}\\

\hline
    \end{tabular}
\tablenotetext{a}{The position angle of the disk plane projected on the sky.}

\end{center}
\end{table}

\section{Discussion}

\subsection{Rotating Disk}

We use three simple geometric models of the ionized gas to
compare to the observed centroid motion of Br$\gamma$ emission:
the gas at the inner rim of a rotating disk, in an accretion
flow, and in a disk wind.

\begin{figure*}
\begin{center}
\includegraphics[height=0.62\textheight,angle=-90]{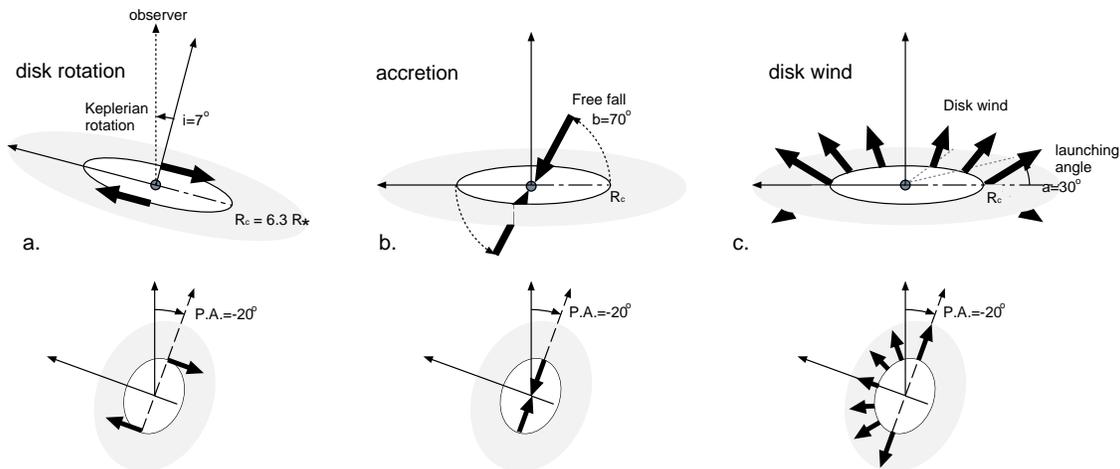}
\end{center}
\caption{Schematics of the geometry assumed in the calculations
  of the model astrometry shown in Figure~\ref{f6}. The figures
  on the bottoms are the model disks on the top row, but in
  projection on the sky with the proper position angle of the
  disk plane. \label{f5}}
\end{figure*}

In the rotating disk model, the plasma is at the inner edge of a
thin disk (Figure~\ref{f5}(a)). The gas is assumed to be in pure
Keplerian rotation with the apparent radial velocity given by $
v_\theta = \sqrt{G M_\ast / R_{\rm c}} \cos{\theta} \sin{(-i)}$,
where $i$ and $\theta$ are the inclination and the azimuthal
angle on the disk plane, respectively. The geometry of the disk
around TW~Hya is reasonably well known.  The disk is almost
face-on ($i<$4\degr), according to the imaging in the scattered
light by {\it Hubble Space Telescope}/WFPC2 and NICMOS
\citep{kri00,wei02}. \citet{pon08} performed 1D
spectroastrometry on CO fundamental line and proposed similarly
small inclination angle ($i=4$\degr$\pm1$\degr). We use
$i=7$\degr~reported in the submillimeter CO observation by the
SMA \citep{qi04,qi06} in the following models, which is
constrained in spatially and spectroscopically resolved
manner. The position angle of the disk major axis is set to
P.A.$=$$-$20\degr, equal to the position angle of the loop
observed in the present study. The stellar radius
($R_\ast=0.96~R_\odot$) and the mass of the star
($M_\ast=0.8~M_\odot$) are taken from \citet{yan05} and
\citet{cal02}, respectively. The disk is assumed to be truncated
inward at the co-rotational radius $R_{\rm c} = 6.3~R_\ast$
\citep[][]{joh01}. The stellar and disk parameters used in the
models are summarized in Table~\ref{tb1}.

The centroid of Br$\gamma$ emission in this model is displaced
to the north at the blueshifted wing, and moves across the disk
with the wavelength along the position angle perpendicular to
the disk rotation axis. This is qualitatively consistent with
the observation (Figure~\ref{f6}(a)). However, the apparent radial
velocity of the gas is too slow to match the observation.
Keplerian velocity at the co-rotation radius is $v \sin{i} <$
19~km~s$^{-1}$. After convolved with the thermal broadening ($v
=$9~km~s$^{-1}$ for 10$^4$K) and the instrumental resolution ($v
=$75~km~s$^{-1}$), the line width of Br$\gamma$ emission is
still too narrow by itself to explain the observed line width of
150~km~s$^{-1}$ in FWHM (Figure~\ref{f6}(a)).

\subsection{Accretion Inflow}

In the second model, the ionized gas is in a pair of accretion
columns starting on a sphere at the distance of the co-rotation
radius. The gas free falls onto the star in straight lines
symmetrically placed about the central star (Figure~\ref{f5}(b)).
The radial velocity component of the free-falling gas is $ v_r =
\sqrt{ 2 G M_\ast ( 1/r - 1/R_\ast)} \cos{(i + b)}$ at the
distance of $r$ with the impact angle $b$ from the disk plane.
The misalignment between the accretion columns and the disk
rotation axis is assumed to be small, with $b=$ 70\degr, so that
the maximum radial velocity matches to the broad wings of
Br$\gamma$ emission ($\pm$200~km~s$^{-1}$).
The azimuthal angle of the inflow projected on the disk is set
to $-$20\degr, so that the apparent position angle of the
astrometry is consistent with the observations.

\begin{figure*}
\begin{center}
\includegraphics[height=0.46\textheight,angle=0]{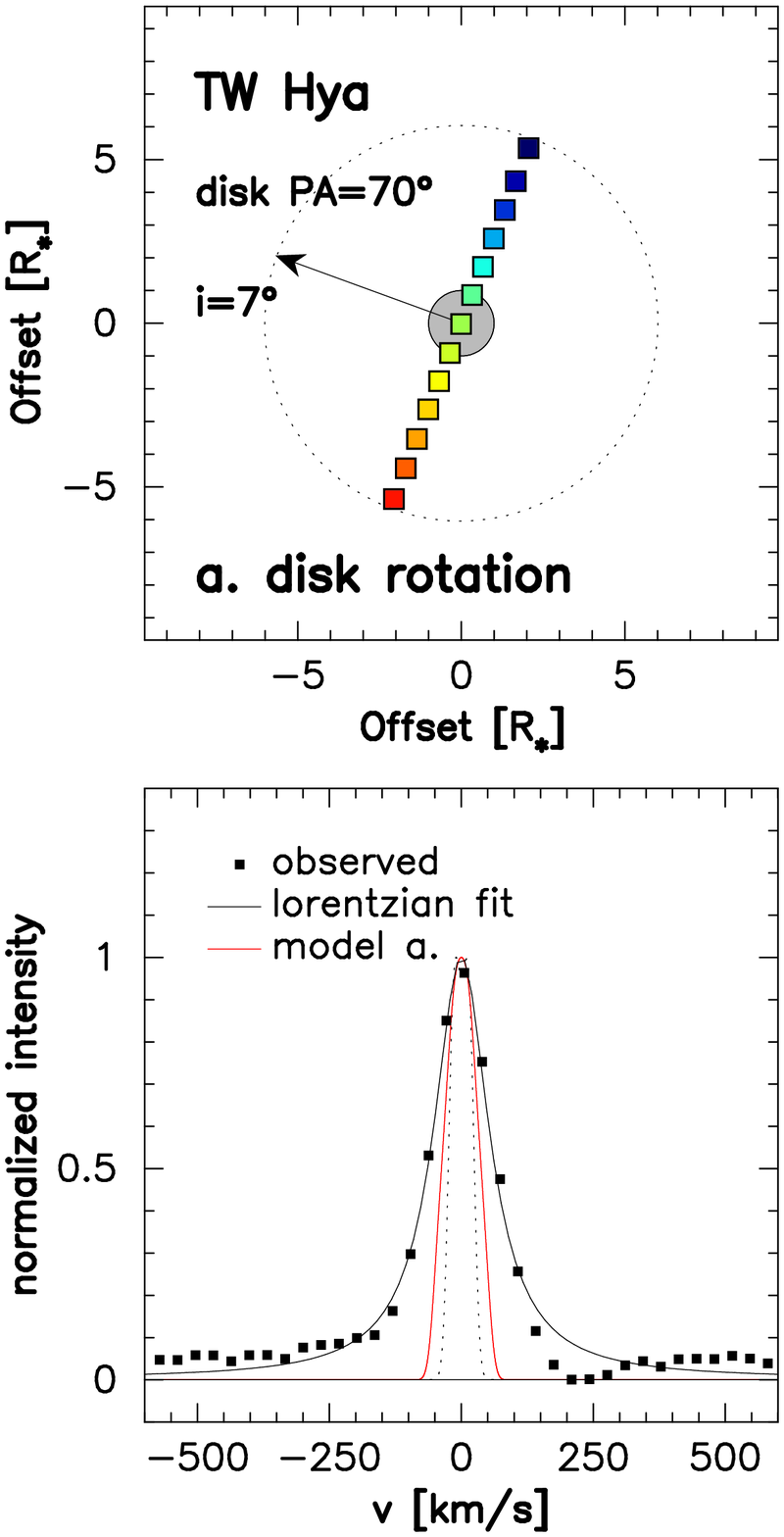}
\includegraphics[height=0.46\textheight,angle=0]{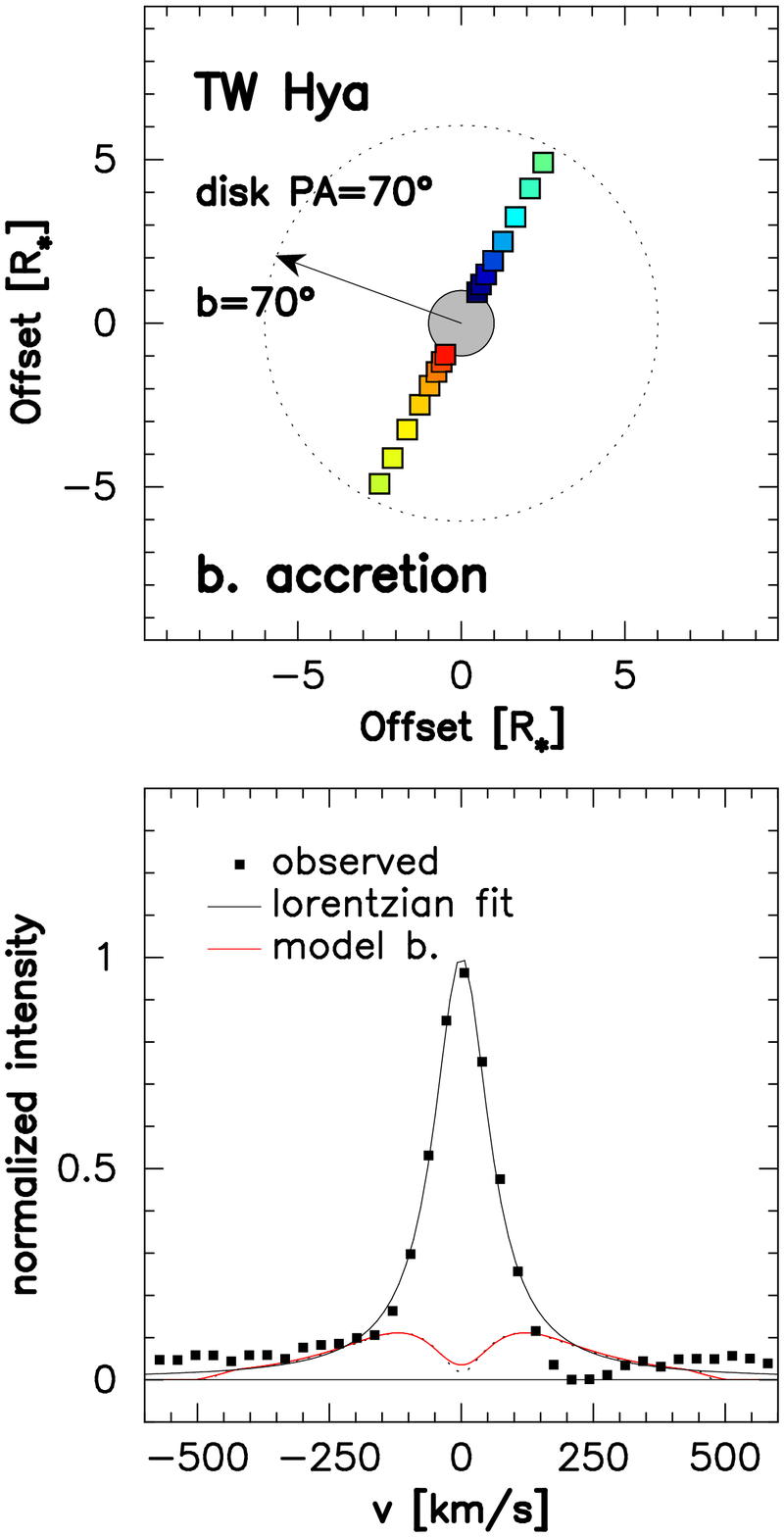}
\includegraphics[height=0.46\textheight,angle=0]{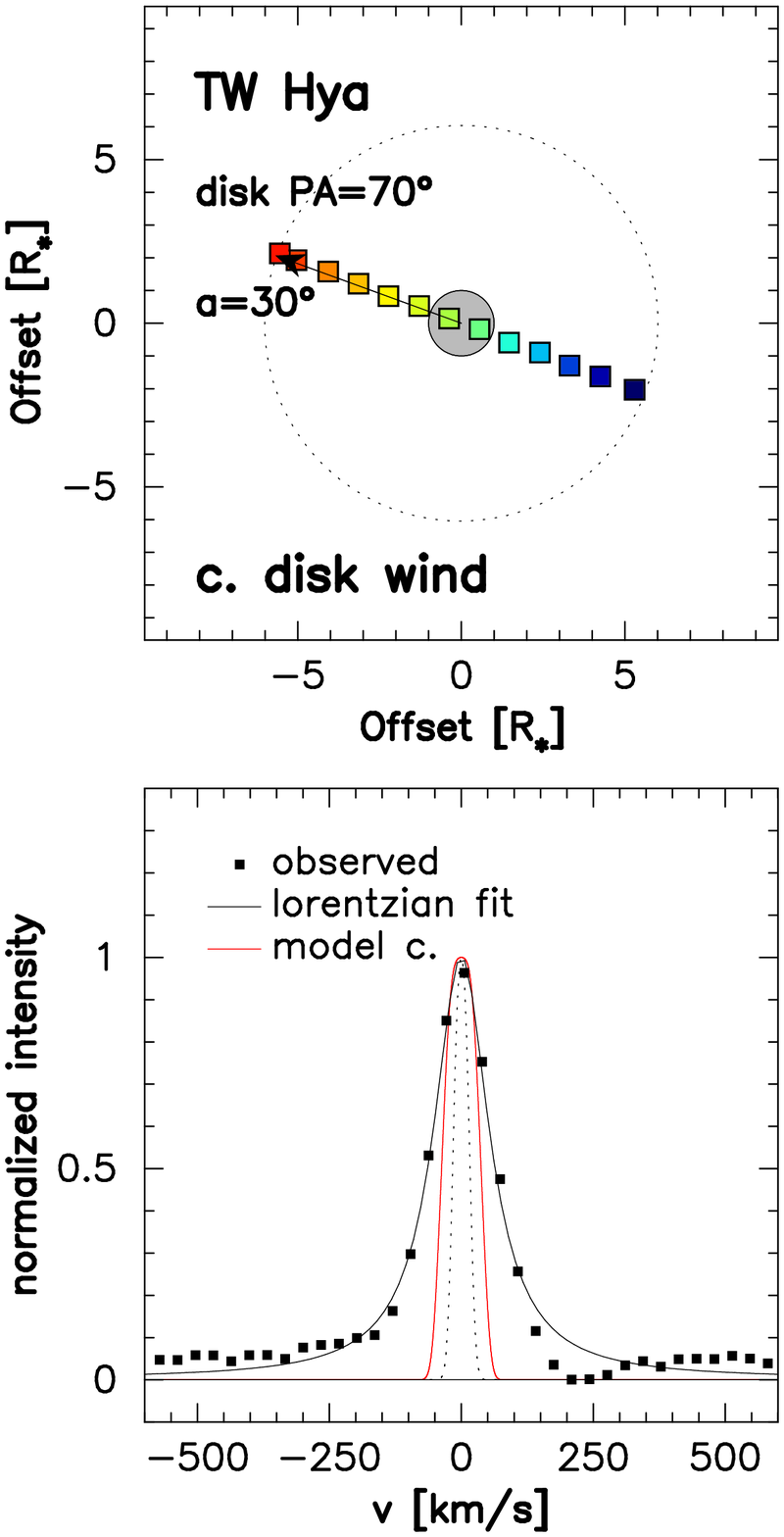}
\end{center}

\caption{Models of two-dimensional astrometry for 
disk rotation (left), accretion flow (middle), and disk winds (right). The upper panels show the
expected centroid motion of Br$\gamma$ emission. The
co-rotational radius $R_{\rm c} = 6.3~R_\ast$ \citep{joh01},
where the disk rotation, accretion flow, or disk winds set
off, is shown by dotted line. The filled circles at the center
represent the actual size of the star
\citep[$R_\ast=0.96~R_\odot$; ][]{yan05}. The lower panels 
show the observed line profile in dots, a Lorentzian fit to the observation in 
black line, and  the line profiles that come out of the models in red. The assumed geometries of the ionized gas are illustrated in Figure~\ref{f5}.
\label{f6}}
\end{figure*}

Br$\gamma$ emission appears most blueshifted at the central
star, as the gas in free-fall is accelerated until it hits the
stellar surface. The centroid of the line emission moves
outward with the wavelength along the accretion column up to
the co-rotational radius. The line emission then jumps to the
other end of the co-rotational circle to the near-side accretion
column at the red half of the emission line
(Figure~\ref{f6}(b)). The apparent motion of the line centroid is
qualitatively consistent with the observation.  However, it
remains unexplained how the position angle of the accretion
inflow coincides with that of the disk major axis at the time of
the observation. The disk of TW~Hya is almost pole on. If the
accretion columns are a single pair of stream lines tied to the
stellar surface by the magnetic field, the position angle
projected on the sky can be of any direction at a given moment,
as it rotates with the stellar rotation.

\subsection{Disk Wind}

In the third model, the disk wind launches outward at the
co-rotation radius with the launching angle $a =$
30\degr~(Figure~\ref{f5}(c)).  The launching velocity is
arbitrarily set to $v_0 =$100~km~s$^{-1}$.  The radial component
of the disk wind at the disk azimuthal angle $\theta$ is $
v_\theta = -v_0 \sin{a} \cos{i} - v_0 \cos{a} \sin{(i)}
\sin{(\theta)}$ at the near-side of the disk.

The centroid of Br$\gamma$ emission at the bluest wing appears
at the intersection of the co-rotational circle and the disk
rotation axis projected on the sky (Figure~\ref{f6}(c)). The
centroids of the line emission moves along the disk rotational
axis with the wavelength from one end to the other end of the
co-rotation circle. The orientation of the motion is
perpendicular to what is observed. Disk winds should only have a
limited contribution to Br$\gamma$ kinematics on the physical
scale we discuss here.

\section{Concluding Remark}

We conclude that the observed astrometry of the Br$\gamma$
emission can be accounted for either by disk rotation or
accretion inflow. The disk rotation, however, does not explain
the line width of Br$\gamma$ emission by itself (150~km~s$^{-1}$
in FWHM), and the accretion inflow requires arbitrary
assumption that the position angle of the accretion columns are
aligned to the major axis of the disk ellipse at the time of the
observation. No clear evidence for disk winds is found.

It is of interest to monitor TW~Hya with the spectroastrometric
technique over its full rotational periods.
TW~Hya is seen almost face-on. In the case that the alignment of
the accretion columns to the position angle of the major axis of
the disk is a sheer coincidence, we should see the locus of the
astrometry rotating on the sky with the stellar rotation.
TW~Hya is the first young star--disk system where a planetary
mass companion might have been found \citep{set08}, although its
presence is questioned by the recent radial velocity
measurements in the infrared \citep{hue08,fig10}. The rotation
period of TW~Hya, inferred by the photometry and the line
emission variability, differs in the literature, ranging from
1.3 to 4.4 days \citep{siw11,mek98,dup07,bat02}.  Ill-determined
rotational period critically hinders clear isolation of the
radial velocity signal of the companion from the stellar
activity.  In the case that Br$\gamma$ emission traces the
accretion columns that are directly connected to the star,
monitoring observation with 2D spectroastrometry will help to
visualize the stellar rotation and to provide the unambiguous
rotational period. The present observations were performed on
the two separate nights, on 2006 January 19 (MJD 53754.20) and
2006 January 8 (MJD 53763.27). The interval was unfortunately
too close to the multiples of the periods reported to date, and
leaves little clues as of the stellar rotation.

\acknowledgments {We appreciate the constructive criticisms of
  the anonymous referees that substantially improved the
  manuscript.  We thank all the staff and crew of the VLT for
  their valuable assistance in obtaining the data. M.G. thanks
  Sebastian Egner who pointed out that the calculation of
  astrometric accuracy is same with that of wavefront error in
  Shack-Hartmann sensor. M.G. was supported by a Japan Society
  for the Promotion of Science fellowship.}


\end{document}